\documentclass[amsmath, amssymb, preprintnumbers, showpacs, showkeys,aps,pra,superscriptaddress,twocolumn]{revtex4-1}

\usepackage{color} 
\usepackage{hyperref}
\usepackage{slashed}
\usepackage{graphicx}
\usepackage{latexsym}
\usepackage{epstopdf}
\usepackage{amsmath}
\usepackage{amssymb}
\usepackage{enumitem}
\usepackage{multirow}
\usepackage{mathtools}
\usepackage{empheq}
\usepackage{bbm}
\usepackage{bm}
\usepackage{amsfonts}
\usepackage{calligra}
\usepackage{calrsfs}
\usepackage{makecell}
\usepackage[normalem]{ulem}
\usepackage{physics}

\begin{document}

\title{Dimensionality-enhanced quantum state transfer in long-range interacting spin systems}

\author{Samihr Hermes}
\email{hermes@fisica.ufrn.br}
\affiliation{Departamento de F\'isica Te\'orica e Experimental, Universidade Federal do Rio Grande do Norte, Natal-RN, Brazil}

\author{Tony J. G. Apollaro}
\affiliation{Department of Physics, University of Malta, Msida MSD 2080, Malta}
\email{tony.apollaro@um.edu.mt}

\author{Simone Paganelli}
\email{simone.paganelli@univaq.it}
\affiliation{Dipartimento di Scienze Fisiche e Chimiche, Universit\`a dell'Aquila, via Vetoio, I-67010 Coppito-L'Aquila, Italy}

\author{Tommaso Macr\`i}
\email{macri@fisica.ufrn.br}
\affiliation{Departamento de F\'isica Te\'orica e Experimental, Universidade Federal do Rio Grande do Norte, Natal-RN, Brazil}
\affiliation{International Institute of Physics, Natal-RN, Brazil}
\affiliation{Dipartimento di Scienze Fisiche e Chimiche, Universit\`a dell'Aquila, via Vetoio, I-67010 Coppito-L'Aquila, Italy}
\date{\today{}}

\begin{abstract}
	In this work we study the single-qubit quantum state transfer in uniform long-range spin XXZ systems in high-dimensional geometries. We consider prototypical long-range spin exchanges that are relevant for experiments in cold atomic platforms: Coulomb, dipolar and van der Waals-like interactions. We find that in all these cases the  fidelity  increases with the dimensionality of the lattice. This can be related to the emergence of a pair of bilocalized states on the sender and receiver site due to the onset of an effective weak-coupling Hamiltonian. The enhancement of the quantum state transfer fidelity is more pronounced both with the increase of the couplings interaction range and in going from a 1D to a 2D lattice.
	Finally, we test our predictions in the presence of temperature-induced disorder introducing a model for the thermal displacement of the lattice sites, considered as a set of local adiabatic oscillators. 
\end{abstract}
\maketitle

\section{Introduction}

Quantum information processing requires suitable transfer protocols for transmitting a quantum state between different parties.  This task results to be nontrivial because of the decoherence induced by the unavoidable interaction with an environment.  
Quantum State Transfer (QST) has been achieved over large distances by spatial transmission of the particle carrying the state (\emph{flying qubit}) \cite{Northup2014,
	Duan2001,
	Ritter2012,Walter2018}. Photons are a natural choice for flying qubits and this strategy has been successfully employed in cavity QED devices. Another option, also suitable for long-distance transmission, is creating an entangled state to be shared between the parties, sender and receiver, in order to implement a teleportation protocol\cite{PhysRevLett.70.1895,Ma2012}.

However, photons are not always the ideal choice to implement  scalable quantum architectures where an efficient short-distance transmission is requested, e.g., in solid state based quantum computers. Here a more desiderable option would be exploiting the natural dynamics of some excitations carrying the quantum information encoded in its quantum state. The most widely investigated model to perform such a task results to be a spin-$\frac{1}{2}$ Hamiltonian, where
the initial state is encoded into one or more spins, accessible to a sender, and it is retrieved, after a certain time, on an equivalent set of spins accessible to a receiver \cite{PhysRevLett.91.207901,Christandl2004a,PhysRevA.71.052315,Christandl2005}. 

Different strategies have been proposed to optimize the QST fidelity:
almost perfect state transfer can be obtained in time-independent uniform chains  \cite{Wojcik2005,CamposVenuti2007a, Banchi2011a},  by 
modulated interactions  \cite{Christandl2004a,Eckert2007},
in disordered chains \cite{Burgarth2005,Almeida2018a}, exploiting the ballistic regime of the excitations  \cite{Osborne2004,DiFranco2008,doi:10.1142/S021974990800392X,Paganelli2006,Christandl2004a,Banchi2010a},  in the regime of weakly  coupled  sender/receiver  \cite{Plenio2005a,Wojcik2005,Paganelli2006,Paganelli2013,doi:10.1142/S021974991750037X,CamposVenuti2007a,chetcuti2019perturbativelyperfect},  creating nearly-resonant edge states introducing strong magnetic fields \cite{Plastina2007,qst1,Paganelli2013}, and  by topological protection \cite{Mei2018,2019arXiv190105157L} (For more detailed reviews  see \cite{Bose2007a,APOLLARO2013}).

Many steps forward have been done, both concerning the QST of  many qubit system \cite{qst2,Sousa2014a,Apollaro2015}, and in the implementation schemes ranging from optomechanical arrays  \cite{DeMoraesNeto2016}, quantum dots  \cite{Farooq2015,Kane1998,Loss1997,DePasquale2005} to ultracold atoms \cite{Volosniev2015a,doi:10.1142/S021974991750037X}.

A lot of effort has been put into accomplishing quantum computing hardware with atom traps 
\cite{VRIJEN2001569,Kielpinski2002,Saffman2016}, since these schemes can be easily mapped into 
a many-body spin-$1/2$  model. In this context, different systems, such as polar molecules,  trapped ions, Rydberg atoms, are characterized by long-range interactions, possibly mediated by long-wavelength modes such as cavity photons, decreasing 
with distance as a power law. Here some of the results obtained for next nearest neighbors interactions do not apply, in particular the estimation of the transmission time becomes more demanding because of the breakdown of the Lieb-Robinson bound \cite{Lieb1972}.  QST has been studied for long-range interacting systems  \cite{Gualdi2008,Avellino2006,Almeida2018a,Eldredge2017,Yao2011} with high fidelity and time speedup. 

In this paper we analyze different long-range spin models and
we show that when the quantum channel is made of a two-(three-)dimensional lattice with uniform coupling, it is possible
 to achieve larger fidelity with respect to the case of a one-dimensional channel with the same Hamiltonian parameters.

In Sec.(\ref{XXZ}) we define the spin model and review the basic features of QST in quantum spin chains with short and long-range couplings. In Sec.(\ref{long-range}) we discuss our results on QST with paradigmatic long-range spin exchange scaling as $1/r^\alpha$ ($\alpha>0$) as the Coulomb interaction ($\alpha=1$), dipolar ($\alpha=3$) and van der Waals ($\alpha=6$), as well as the case of ($\alpha=1/2$) relevant for ion-trap experiments in one, two and three dimensional geometries.
In Sec. (\ref{vacancies}) we analyze the effect of vacancies as a way to improve the QST in these models and interpret the results by looking at the spectrum of effective two-spin models. The relevance of symmetry in the removal of spins in the chain is emphasized.
In Sec.(\ref{temperature}) we discuss finite temperature effects leading to a displacement of the spins with respect to their equilibrium position. 
Finally, in Sec.(\ref{conclusions}) we present our conclusions and propose some extensions of our work. 
	
\begin{figure}[!h]
\centering
\includegraphics[width=0.8\columnwidth]{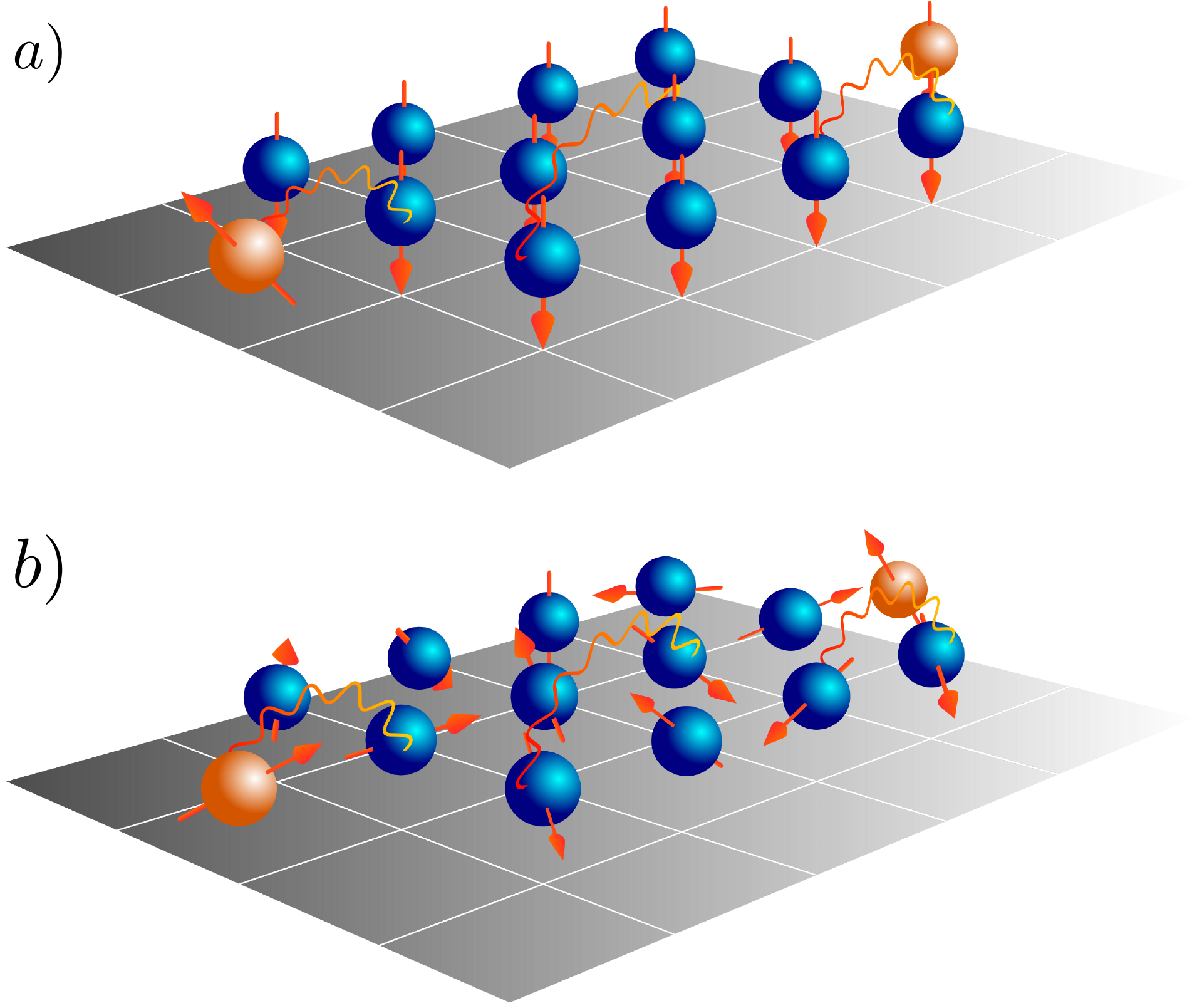}
\caption{Dynamical evolution of a quantum spin system with long-range
interactions in a two-dimensional array. Sender and receiver (orange)
and the channel (blue) interact via a long-range spin-exchange interaction $1/r^{\alpha}$.
At $t=0$ the system is the state $\left|\psi_s\right> =\alpha\left|0\right> + \beta\left|1\right>$ 
with one excitation localized at the sender $i=1$. 
After a certain time $t$ the state $\left|\psi_s\right>$
of the sender cab be found with high fidelity
at the receiver $i=N$.
} \label{fig1}
\end{figure}

\section{Quantum state transfer in the XXZ model.}
\label{XXZ}
We consider a long-range XXZ spin model with Hamiltonian
\begin{equation}\label{Hfirst}
\hat{H} = \sum_{\substack{i,j=1 \\ i <j}}^{N} \frac{C}{2a^{\alpha}\abs{\mathbf{r}_{i}-\mathbf{r}_{j}}^\alpha}\qty(\hat{S}_{i}^{+}\hat{S}_{j}^{-}+\hat{S}_{i}^{-}\hat{S}_{j}^{+}+2\Delta \hat{S}_{i}^{z}\hat{S}_{j}^{z}),
\end{equation}
where $S^\pm$ and $S^z$ are spin-$1/2$ operators. The couplings $C$ and $\Delta$ denote the
intensity of the spin-exchange interaction and the anisotropy parameter, whereas $\alpha$ is the power-law 
exponent the long-range coupling. Finally $a$ is the lattice spacing among nearest neighbor spins in the lattice.
For simplicity we set $C=1$ in units of $\text{Energy}\times a^\alpha$ throughout our work, while ${\bf r_i}$ are dimensionless
positions of the spins in units of $a$.
For $\Delta=0$ one obtains the isotropic long-range XY model, 
for $\Delta=1$ one recovers the  isotropic long-range Heisenberg model, 
and for $\Delta=\infty$ we obtain the long-range (classical) Ising model. The nearest-neighbor ($\alpha=\infty$) 
isotropic Heisenberg model was first considered
in the seminal paper by Bose in \cite{Bose}. See Appendix \ref{nearest-Bose} for the generalization of these results to higher dimensions.
In this work we vary the long-range exponent $\alpha>0$ and set the anisotropy parameter $\Delta=-2$. 
We notice that for $\Delta=-2$ we can rewrite the spin couplings in the form of a dipolar exchange potential
\begin{equation}
\hat H_{dip} = \sum_{ij}\, J_{ij} \left({\bf \hat S_i} \cdot {\bf \hat S_j} - 3 \hat S_i^z \hat S_j \right),
\end{equation} 
where $J_{ij} = C / 2a^{\alpha}\abs{\mathbf{r}_{i}-\mathbf{r}_{j}}^\alpha$.

The protocol describing the dynamics of our setup is described in Fig.\ref{fig1} for a two-dimensional setup.
The one- and three-dimensional setups will be explicitly discussed below.
In fig.(\ref{fig1})a the system of $N$ spin-$1/2$ is initialized in the ferromagnetic 
state $\otimes_i \ket{\downarrow}_{i}$ in the $z-$basis. 
At $t=0$ one spin, the sender (in orange) is placed into a state  
$\ket{\psi_{s}} = \cos{\frac{\theta}{2}} \ket{\downarrow} + e^{i\phi}\sin{\frac{\theta}{2}}\ket{\uparrow}$. 
Here $\theta \in [0,\pi ]$ and $\phi \in [0,2\pi ]$ are the usual angles defining a single qubit state in the Bloch sphere.
The system is then left to evolve under unitary dynamics with the Hamiltonian eq.(\ref{Hfirst}).
Additional effects such as decoherence, excited state decay, or a generic coupling to an external reservoir, 
will be considered elsewhere, while the effects of temperature-induced positional disorder will be thoroughly discussed in Sec.~\ref{temperature}.

We observe that for the model we are considering the total magnetization is preserved, i.e., $\comm{\hat{H}}{\sum_{i=1}^{N}S^{z}}=0$. Therefore the Hilbert space where the dynamics takes place is confined to the zero-excitation sector
consisting of the fully ferromagnetic state $\otimes_i \ket{\downarrow}^{i}$ and the
one excitation sector consisting of $N$ states, where $N-1$ spins are in the $\ket{\downarrow}$ configuration 
and one is in the $\ket{\uparrow}$ configuration.
Under these conditions one can redefine the many-body states in the computational basis as
\begin{eqnarray}
\otimes_{i} \ket{\downarrow}_i & \equiv &\ket{\mathbf{0}}; \\
\ket{\uparrow}_j \otimes_{i\neq j} \ket{\downarrow}_i & \equiv & \ket{\mathbf{j}},
\label{basis}
\end{eqnarray}
that belong to a subspace $\mathbf{H}$ of dimension $N+1$ of the full Hilbert space.
For our calculations we perform exact numerical diagonalization of the Hamiltonian matrix of eq.(\ref{Hfirst}) in the basis above.
Taking $\frac{C}{2a^{\alpha}} = 1$, we have the diagonal and off-diagonal elements
\begin{align}
\label{E_matrix_ele}
\bra{\mathbf{j}}H\ket{\mathbf{j}} & = \frac{\Delta}{2} \qty( \sum_{\substack{k,l=1 \\ k \neq l}}^{N}
\frac{1}{\abs{\mathbf{r}_{k}-\mathbf{r}_{l}}^\alpha} - \sum_{\substack{i=1 \\ i \neq j}}^{N}
\frac{1}{\abs{\mathbf{r_{i}-\mathbf{r}_{j}}}^\alpha}),\\
\bra{\mathbf{i}}H\ket{\mathbf{j}} & = \frac{1}{\abs{\mathbf{r}_{i}-\mathbf{r}_{j}}^\alpha}.
\end{align}
For the special case of nearest neighbor exchange interactions one has
\begin{align*}\label{LRmatrix_elements2DNN}
\bra{\mathbf{i}}H\ket{\mathbf{j}} & = 
	\begin{cases}
		   0 \quad \text{if} \quad \abs{\mathbf{r}_{i}-\mathbf{r}_{j}} \neq 1 \, \text{;}\\
		   1 \quad \text{otherwise.}
	\end{cases} \\
\bra{\mathbf{j}}H\ket{\mathbf{j}} & = \sum_{\substack{i,k=1 \\ i<k}}^{N} S_{ik}.
	\end{align*}
where
\begin{equation}
S_{ik} = \begin{cases}
		   \, \, \, \, \frac{\Delta}{2} \, \, \text{if} \, \, \abs{\mathbf{r}_{i}-\mathbf{r}_{k}}=1 \, \, \text{and} \, \, j \neq i,k \, \text{;}\\
		   \, \, \, \, \, -\frac{\Delta}{2} \, \, \text{if} \, \, \abs{\mathbf{r}_{i}-\mathbf{r}_{k}}=1 \, \, \text{and} \, \, j=i \, \text{or} \, j=k \, \text{;}\\
		   \, \, \, \, \, 0 \, \, \text{otherwise.}
		   \end{cases}
\end{equation}

As a figure of merit of the quality of the QST 
from the sender to the receiver (orange spins in fig.(\ref{fig1})) we use the  fidelity 
\begin{equation}
F \qty(t) = \int \frac{\dd{\Omega}}{4\pi} \bra{\psi_s\qty(t)}\rho_r(t)\ket{\psi_s\qty(t)}~
\end{equation}
where $\rho_r(t)$ is the density matrix of the spin at the receiver site $N$ and the 
average is taken over the initial state of the sender.
 
Upon integration we obtain the general expression
\begin{equation}\label{Fidelity}
F\qty(t) = \frac{1}{6}\abs{f_{r,s}\qty(t)}^2 + \frac{1}{3}\abs{f_{r,s}\qty(t)}+\frac{1}{2}.
\end{equation}
where we defined $f_{r,s}\qty(t) = |\bra{\bf{r}}e^{-i \hat H t} \ket{\bf{s}}|$ and set $\hbar=1$. 
$\ket{\bf{s}}$ and  $\ket{\bf{r}}$ are the singly-excited states localized at the sender and receiver respectively.
In the calculation of $F(t)$ we also neglected the term $\cos(\gamma)$, where $\gamma$ is the phase of the 
amplitude $\bra{\bf{r}}e^{-i \hat H t} \ket{\bf{s}}$.

In the next paragraphs we analyze the dynamics of $F(t)$ for configurations in one, two and three dimensions
as a function of the power-law exponent $\alpha$.

\section{Quantum state transfer with long-range couplings.}
\label{long-range}

\begin{figure}[!h]
\centering
\includegraphics[width=1.0\columnwidth]{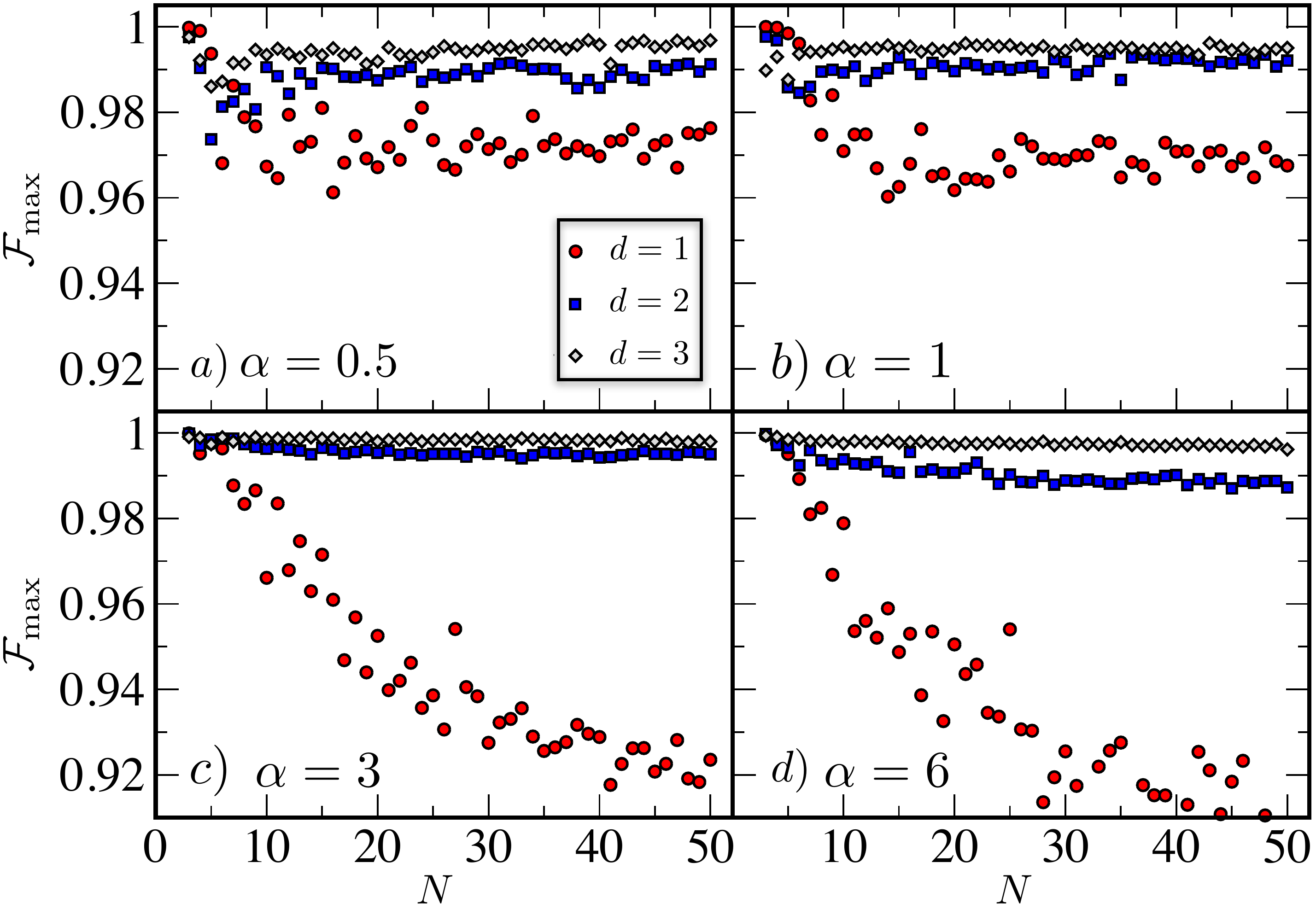}
\caption{Maximum fidelity for quantum state transfer in a long-range interacting system in one, two and 
three dimensions for different power-law interaction potentials and $\Delta=-2$. 
Red dot: $d=1$, blue square: $d=2$, grey diamond: $d=3$.
$N$ is the linear size of the system. 
In 2D (3D) we consider a rectangle (parallelepiped) lattice with $N\times 5$ ($N\times 5 \times 5$) spins.
In 2D the sender and the receiver are located as in Fig.\ref{fig1}. In 3D they are located in the center
of the square-section of opposite faces of the parallelepiped, at the distance $a$ from the central spin.
The exponent of the power-law interaction is a) $\alpha=0.5$, b) $\alpha=1$, c) $\alpha=3$ and d) $\alpha=6$.
Fidelity increases with the the dimensionality of the system for each power-law interaction. 
We notice that in 3D the fidelity stays close to one even for a large system size.
}\label{fig2}
\end{figure}

In this Section we discuss the maximum fidelity achievable for the QST in a long-range interacting system in one, two and 
three dimensions for different power-law interaction potentials. In fig.(\ref{fig2}) we show the results of the simulations. 
In the one dimensional configuration (red dots) sender and receiver are located at the extremes of the the chain. 
In 2D (blues squares) we consider a rectangle lattice with $N\times 5$ spins where sender and receiver 
are placed as in fig.(\ref{fig1}). In 3D (grey diamonds) we examine a cubic lattice with ($N\times 5 \times 5$) spins where
sender and receiver are located in the center of two opposite faces of the parallelepiped
with $5\times 5$ spins, at the distance $a$ from the central spin.
The exponent of the power-law interaction that we analyze are a) $\alpha=0.5$, b) $\alpha=1$, c) $\alpha=3$, and d) $\alpha=6$.
We notice that the higher the dimensionality of the lattice and the range of the interaction, the higher is the fidelity. Furthermore, the enhancement is more pronounced the larger the system and, in 3D lattices, the fidelity even stays close to one for relatively large system sizes. 

The dimensionality and interaction-range enhanced fidelity can be traced back to the onset of an effective weak-coupling regime taking place because of the open boundary conditions of the lattice at whose edges the sender and receiver spin are coupled symmetrically. Protocols exploiting the weak-coupling regime between sender (receiver) and a quantum channel for QST have been widely explored mainly in one-dimensional lattices. In 1D and in the presence of short-range Hamiltonian, to obtain the Rabi-like oscillations of the excitation between the sender and the receiver site, it is necessary effectively decouple the sender and the receiver site from the quantum wire. This can be obtained both by acting on the respective hopping term and on the on-site magnetic field in the Hamiltonian, or on both at the same time. Here we consider a Hamiltonian with uniform coupling and the onset of an effective weak-couplings regime is a combined effect of the lattice dimensionality, the range of the interaction and the presence of the spin-exchange interaction term in the transverse $z$ direction. The effect of the latter on the  Hamiltonian in the single-excitation sector mimics that of an effective non-homogeneous on-site magnetic field in the $z$-direction with values given by Eq.~\ref{E_matrix_ele}.
 
To explain in a quantitative way the increase of the maximum of the fidelity with the dimensionality, we analyze the 
eigenvectors of the Hamiltonian in eq.(\ref{Hfirst}) that have maximum overlap with the sender and the receiver 
We observe that for the topology we discuss in this work by increasing the dimensionality of the lattice, the connectivity 
of the sender and the receiver with the neighbor spins of the channel (with strongest coupling) decreases with respect to the connectivity of the channel spins in the bulk. 
The ratio of the connectivities of the sender/receiver with the channel spins equals $C_{s/r}/C_c \, = 1/2,\, 1/4,\, 1/6$
respectively in one, two, and three dimensions. 
Therefore, the eigenstates localize more on the sender and receiver spin in higher dimensions, leading to a higher fidelity of the quantum state transfer.

\subsection{Quantum state transfer in the presence of defects and the role of symmetry.}
\label{vacancies}

\begin{figure}[!h]
\centering
\includegraphics[width=1.0\columnwidth]{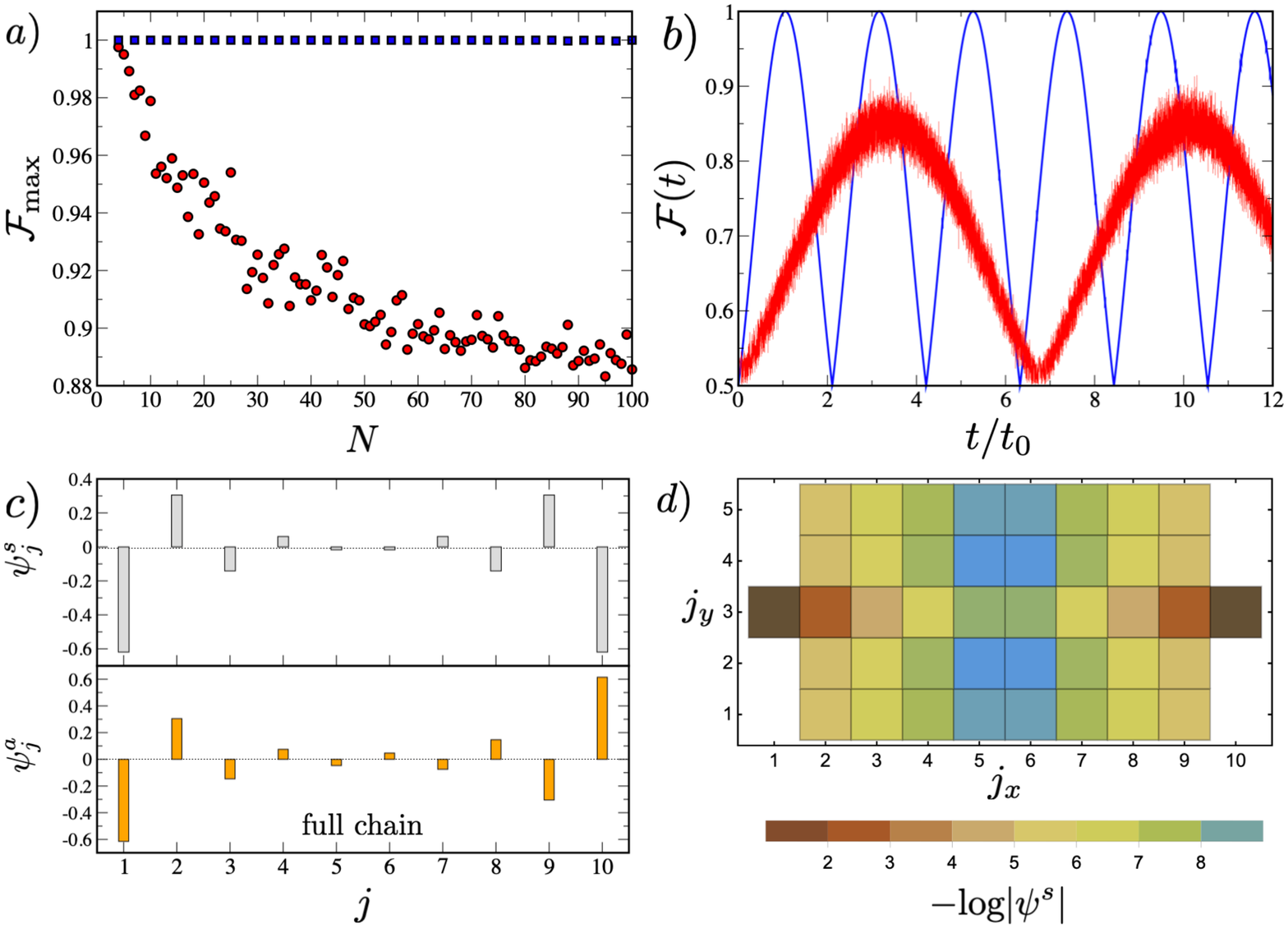}
\caption{Quantum state transfer for a long-range interacting one- and two-dimensional system with $\alpha=6$ 
(van der Walls interactions) and $\Delta=-2$. 
a) Maximum value of the fidelity for a system with $N$ spins ($N=2,\dots,100$).
Red dots: One-dimensional chain with $N$ spins.  
Blue squares: Two-dimensional $N\times 5$ setup as in d).
b) Dynamical evolution of the fidelity as a function of the rescaled time $t/t_0$. The time
$t_0= \pi a^\alpha (N-1)^\alpha/C \hbar^2$ is the time it takes for the 
two-spin system (sender and receiver) to perform an ideal quantum state transfer for $N=50$.
c) Symmetric and antisymmetric eigenstates $\psi^{(s,a)}$
with the greatest overlap with the sender and the receiver, which are responsible
for the high-fidelity state transfer for the one-dimensional case.
d) Overlap with highest overlap with the sender and the receiver for a two-dimensional system with $10\times 5$ spins. Darker colors correspond to higher overlap.
}\label{fig3}
\end{figure}

In fig.(\ref{fig3})a we plot the maximum value of the fidelity for a system with $N$ spins ($N=2,\dots,100$)
with the one-dimensional chain (red dots) and two-dimensional lattice with $10\times 5$ spins (blue squares).
In fig.(\ref{fig3})b we show the dynamical evolution of the fidelity as a function of the rescaled time $t/t_0$, where
we define $t_0= \pi a^\alpha (N-1)^\alpha/C \hbar^2$. 
$t_0$ is the time it takes for the two-spin system (sender and receiver) to perform an ideal quantum state transfer in the absence of the channel.
As we discussed in the previous paragraph, a quantitative explanation of the increase of the fidelity can be obtained by
studying the eigenstates with maximal overlap with the sender and the receiver.
In fig.{\ref{fig3}}c-d we plot the symmetric and antisymmetric eigenstates $\psi^{(s,a)}$ 
which are responsible for the state transfer for the regular chain with $N$ spins (c) and for the two-dimensional lattice $10\times 5$ (d).
The effect is an increase of the fidelity and a modification of the period of the oscillations of the fidelity.

\begin{figure}[!t]
\centering
\includegraphics[width=1.0\columnwidth]{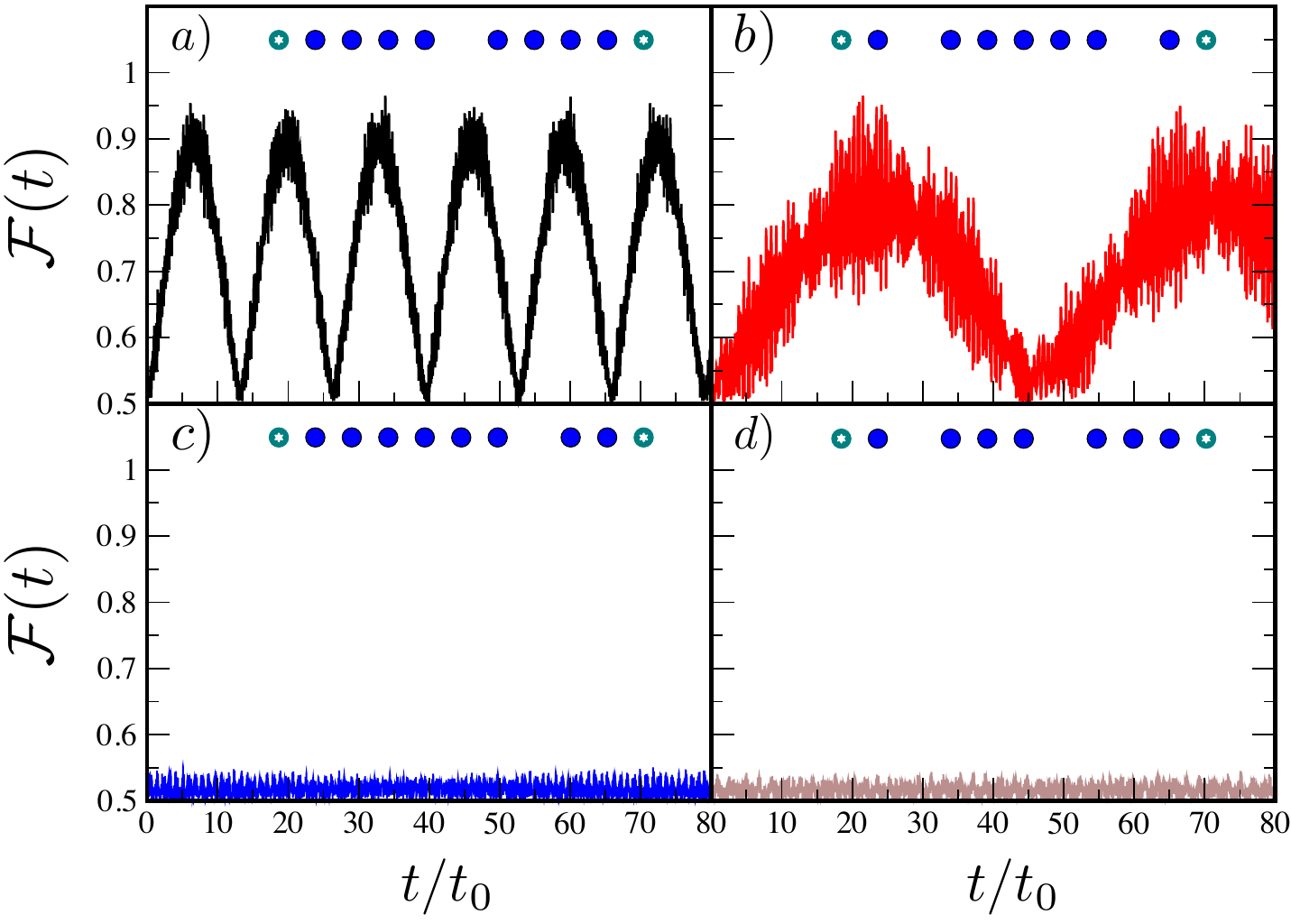}
\caption{Quantum state transfer and mirror symmetry of the vacancies 
in a linear chain with $\alpha=1$.
Dynamical evolution of the fidelity $\mathcal{F}(t)$ with one vacancy a) and c), and 
two vacancies b) and d). 
In a)-b) spins are removed symmetrically with respect to the center of the
chain. In c)-d) vacancies are created randomly along the chain. 
}\label{fig4}
\end{figure}

\begin{figure}[!t]
\centering
\includegraphics[width=1.0\columnwidth]{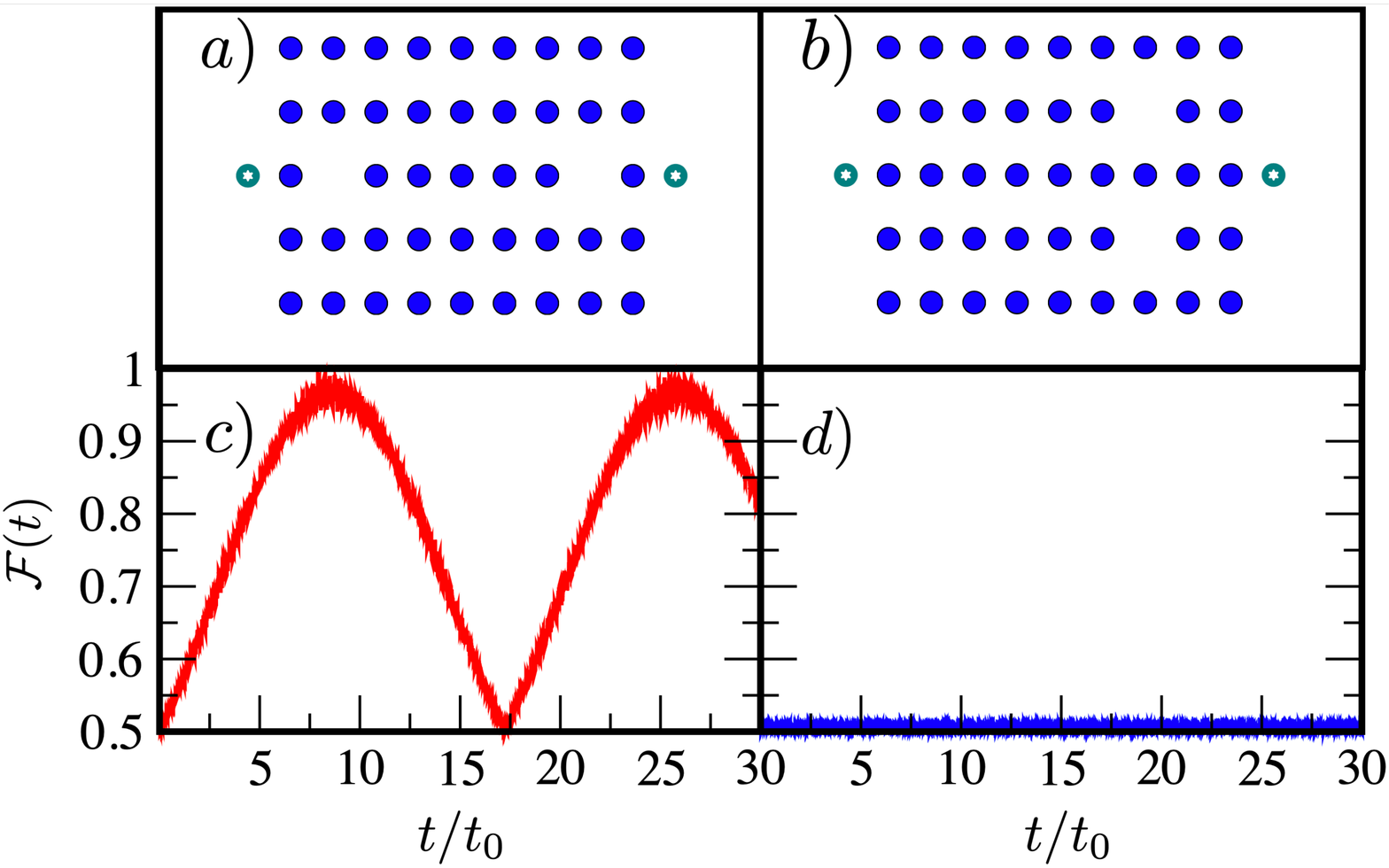}
\caption{Quantum state transfer and mirror symmetry of the vacancies 
in a 2D square lattice with $\alpha=1$.
Dynamical evolution of the fidelity $\mathcal{F}(t)$ with a) two symmetric vacancies and c) 
two asymmetric vacancies. 
}\label{fig5}
\end{figure}

The Hamiltonian we consider in our work is mirror symmetric, which in Refs.~\cite{Christandl2004a, Christandl2005} was found to be a necessary condition for perfect QST.
Here we investigate the breaking of the mirror symmetry by 
removing arbitrary spins in the configurations. For simplicity we focus on one- and
two-dimensional systems. The results are reported in fig.(\ref{fig4}) and fig.(\ref{fig5}).

In fig.(\ref{fig4}) we plot the configuration and the dynamics of the QST
fidelity 
in a linear chain with Coulomb interaction $\alpha=1$ in the presence of one and two 
vacancies.
In a)-b) spins are removed symmetrically with respect to the center of the
chain. 
In c)-d) vacancies are created randomly along the chain.  
Without mirror symmetry the fidelity decreases to approximately the random guess value of $\frac{1}{2}$. This corresponds to a vanishing probability for the excitation to reach the sender site, see. Eq.~\ref{Fidelity}. However, when in the presence of mirror symmetry, the fidelity has maxima close to $1$.

We repeat the analysis for a two-dimensional system with a channel with 
$9\times 5$ spins. We observe that, preserving mirror symmetry as in fig.(\ref{fig5})a-b, 
the dynamics displays a high value of the fidelity. In contrast, when mirror symmetry is broken by the removal of two spins, fidelity decreases again to around $\frac{1}{2}$.

\subsection{Finite temperature effects.}
\label{temperature}

\begin{figure*}[!t]
\centering
\includegraphics[width=2.0\columnwidth]{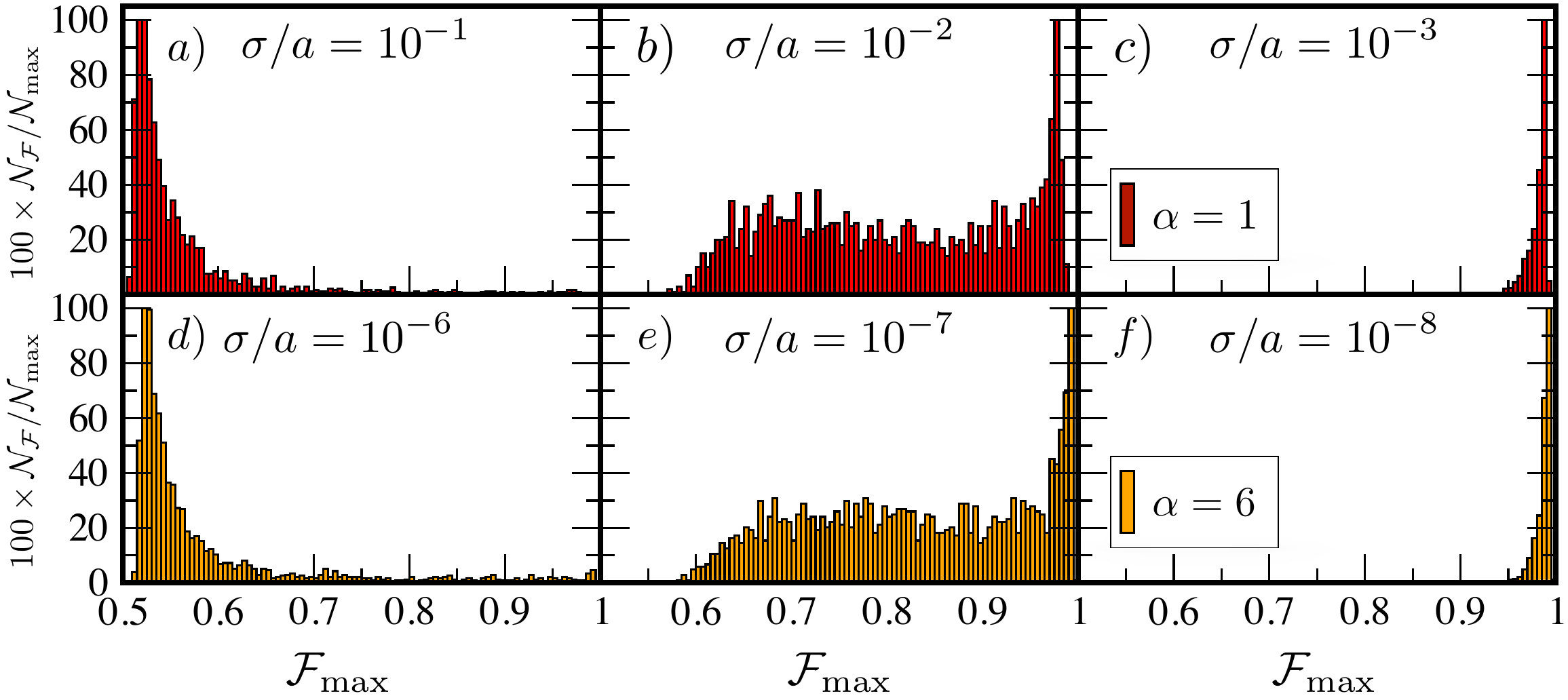}
\caption{Distribution of the maxima of the fidelities for gaussian disordered 
atomic position configurations for several values of the disorder strength. 
Particles are displaced with respect to their equilibrium value according to
a gaussian distribution with width $\sigma/a$ in a 2D lattice with length 
$L=10$ and height $h=4$.
The strength of the disorder is 
connected to the temperature of the configuration as explained in the text.
We consider $N_\text{r}=2000$ realizations.
a)-c) $\alpha=1$ Coulomb spin-exchange interaction.
d)-f)  $\alpha=6$ van der Walls spin-exchange interaction.
The distribution is normalized to the
peak value of each histogram.}
\label{fig6}
\end{figure*}

We now discuss the effect of disorder in the particle configuration. 
We focus on the two-dimensional case. However the results can be 
generalized to both one and three dimensions straightforwardly. 
The motivation for this analysis is related to recent experiments on Rydberg atoms
trapped in optical tweezers. There atoms are trapped in a strongly focussed 
laser field, with a small but finite dispersion of the position. 
We model this effect as a temperature-induced quenched disorder 
on the particle configuration as in \cite{Barredo2018}.
This description is valid if the dynamics of the spin is decoupled from the 
motional degrees of freedom, i.e. if the time $t_{id}$ associated to the QST protocol is much faster than the typical motional time scales $\hbar/k_B T$,
where $T$ is the temperature of the system.
Then, we model the dynamics of the motional degrees of freedom of particle $i$ centered
in the lattice site with coordinates $(x^0_i,y^0_i)$ with a classical 
Boltzmann distribution 
$f(\mathbf{r},\mathbf{p}) = \exp \qty[-\beta H_{m}\qty(\mathbf{r},\mathbf{p}))]$, where
\begin{equation}
H_{m}^{(i)}\qty(\mathbf{r},\mathbf{p}) = \sum_{j=x,y} 
\frac{p_{j}^2}{2m} + \frac{m}{2} \omega_{j}^2(r_{j}-r_j^0)^2.
\end{equation}

To find the distribution of the position $\bar f(x,y)$, we integrate the momentum contribution and normalize
\begin{equation}\label{posdist}
\bar f(x,y) = \frac{1}{2\pi} \frac{1}{\sigma_{x}\sigma_{y}} \exp\qty[-\frac{\qty(x-\bar{x})^2}{2\sigma_{x}^2}-\frac{\qty(y-\bar{y})^2}{2\sigma_{y}^2}],
\end{equation}
where we defined the variance $\sigma_i^2 = 1/(\beta m \omega_i^2)$.
We notice that, although we restrict our analysis to fluctuations along the plane $x-y$, if we were to fully model
an experimental setup, also the confinement along the $z$ direction should be considered. Therefore our study 
corresponds to the limiting case of vanishing $\sigma_z$.
For simplicity, we are restricting to $\sigma_{x}=\sigma_{y}=\sigma$. 
We observe that, although particle positions in the lattice are uncorrelated, 
interparticle distances (in the definition of the spin exchange couplings $J_{ij}$) are correlated \cite{Marcuzzi2017}.

The results of the simulations are shown in fig.(\ref{fig6}).
We plot the distribution of the maxima of the fidelities for gaussian disordered 
atomic position configurations for several values of the disorder strength $\sigma/a$ rescaled to the lattice spacing. 
Particles are displaced with respect to their equilibrium value  
in a 2D lattice with length $L=10$ and height $h=4$.
We consider $N_\text{r}=2000$ realizations of the disorder and $\alpha=1$ Coulomb spin-exchange interaction (a-c) 
and $\alpha=6$ van der Walls spin-exchange interaction (d-f).
For clarity the distribution is normalized to the peak value of each histogram.
Fixing $\alpha$ and for low disorder the distribution is peaked close to unitary fidelities. 
Increasing the disorder strength a plateau appears in the distribution with a peak at $F_{max}=1$.
By further increasing the disorder the distribution has a peak at $F_{max}=1/2$ and 
the plateau disappears. We notice that this behavior is quite generic for the long-range exponents that we
analysed. 
The second relevant feature of the fidelity distributions is that, upon decreasing $\alpha$ smaller values of 
the disorder $\sigma/a$ are needed to obtain a distribution peaked at higher fidelities. 
The qualitative explanation is that a longer-range interaction makes the system more insensitive to
the fluctuation of particle positions. To be more quantitative, this can be seen by computing the variation of the
spin-exchange couplings as a function of the disorder \cite{menu2019anomalous}
\begin{equation}
\frac{\Delta J_{ij}}{J_{ij}} = \frac{\tilde J_{ij} - J_{ij}}{J_{ij}} = \frac{r_{ij}\,^\alpha}{\left|\bf{r}_{ij}+\bf{\delta}\right|^\alpha}-1, 
\end{equation}
where we defined $\bf{r}_{ij}=\bf{r}_i-\bf{r}_j$, $\tilde J_{ij}= C/2a^\alpha \left|\bf{r}_{ij}+\bf{\delta}\right|^\alpha$,
and $\bf{\delta}$ is the difference of the fluctuations of the two particles.
Monitoring the variation of $\frac{\Delta J_{ij}}{J_{ij}}$ as a function of $\alpha$ for a fixed equilibrium 
interparticle spacing and disorder strength, the ratio vanishes for $\alpha=0$. 
This corresponds to a position-independent spin coupling that 
clearly should not depend on the specific value of disorder.
In the opposite limit, when $\alpha$ increases the ratio $\frac{\Delta J_{ij}}{J_{ij}}$ decreases to the limiting value $-1$
for infinite $\alpha$ (position dependent nearest neighbor interaction). 
In this limit disorder dominates and the regularity of the particle configuration, 
including a strong breaking of mirror symmetry, leads to a dramatic reduction of the fidelity of the quantum state transfer.


\section{Conclusions}\label{conclusions}
In this work we studied the problem of quantum state transfer in lattices with open boundary conditions in one, two and three dimensions
for the anisotropic Heisenberg XXZ model with of power-law couplings with variable exponent $\alpha$ and with a sender and receiver spin symmetrically coupled to its edges.

We first analyzed the case of regular lattices and found that the fidelity increases upon increasing the 
dimensionality of the lattice for sufficiently large system sizes and  the enhancement of the fidelity is more pronounced for systems with long-range interaction.
We interpreted this result as a combined effect of the open boundaries of the lattice and the presence of the inter-spin interaction term in the $z$-direction, resulting in an effective weak-coupling Hamiltonian in the single-excitation sector, although the couplings are all uniform.  
We justified this interpretation by noticing that the quantum state transfer takes place via Rabi-like oscillations involving only two single-particle eigenstates localized on the sender and receiver site, a mechanism that is related to resonant tunneling in effective decoupled models.

We studied also the effect of vacancies both in one and two dimensional lattices,  confirming the necessity of the presence of mirror symmetry in the lattice configuration with the removed spins in order to increase the fidelity of the quantum transfer transfer.
Also, for the relevant case of one-dimensional systems, 
we observed that for longer-range interactions, one might consider a larger number of vacancies
to obtain a higher fidelity.
Finally, inspired by experiments of cold atoms in optical tweezers, we considered the effect of a finite
temperature inducing displacements of the particles by studying the distribution of the maxima of the
fidelity.
We observed that, as a general property, longer range interactions suffer less from temperature-induced
disorder than shorter range potentials. Quantitatively, this effect can be understood by analyzing the
fluctuations of the spin couplings $J_{ij}$ due to disorder as a function of the exponent $\alpha$.

Our study is relevant for the characterization of quantum state transfer in experimental
platforms for quantum simulation and technology. 
We were mostly inspired by applications
to ultracold ions and atoms where long-range interaction are an intrinsic tool in the realization of spin models.
Extensions of this work include the study of the effect of decoherence and excited state decay, relevant for
experimental platforms. Calculations for open systems will be considered elsewhere.

\begin{acknowledgments}
{\it Acknowledgments}.
S.H. acknowledges CNPq for financial support. T.M. acknowledges
CNPq for support through Bolsa de produtividade em Pesquisa n.311079/2015-6.
This work was supported by the Serrapilheira Institute (grant number Serra-1812-27802), 
CAPES-NUFFIC project number 88887.156521/2017-00. T.M.
and thanks the Physics Department of the University of L'Aquila 
for the hospitality where part of the work was done. The authors
acknowledge S.M. Giampaolo for useful discussions.
\end{acknowledgments}

\appendix

\section{Nearest neighbor interactions in higher lattices} \label{nearest-Bose}

In this Appendix we present the results of the maximum fidelity for nearest neighbor interactions in a slab with
linear dimension $L$ and transverse length $L_\perp = 5$ in two dimensions and a parallelepiped 
$L\times L_\perp  = 5\times 5$ in three dimensions.
For purely one dimensional system, our results are equivalent to \cite{PhysRevLett.91.207901}
We observe that, similarly to the long-range case, fidelity is higher for a higher dimensional slab. However,
the maximum fidelity is notably smaller than unity even for the three-dimensional case, in contrast to Fig.\ref{fig2} where for each $\alpha \le 6$ fidelity is close to one even for very large sizes.

\begin{figure}[!h]
\centering
\includegraphics[width=1.0\columnwidth]{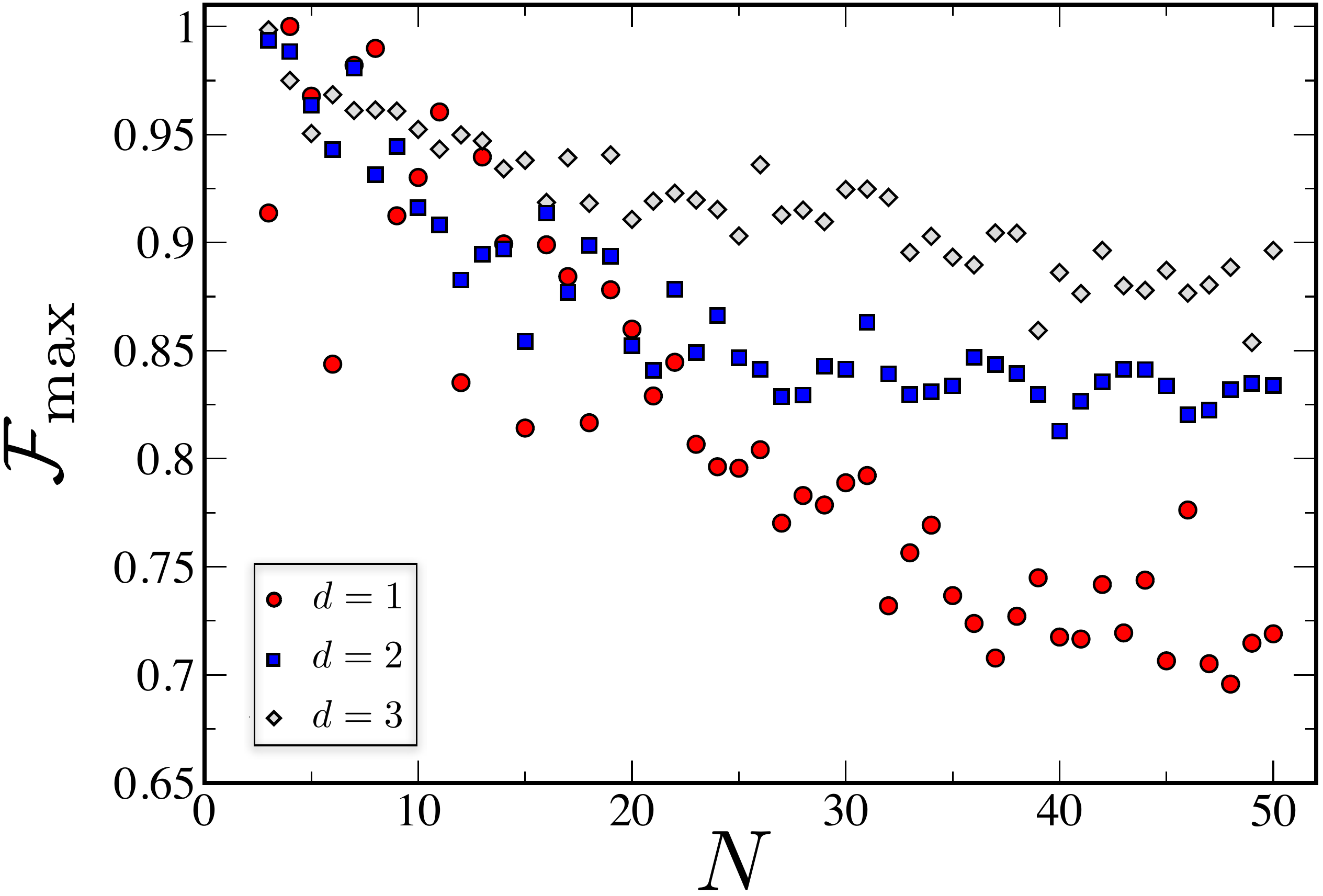}
\caption{Maximum fidelity for quantum state transfer in a nearest neighbor interacting system in a one, two and 
three dimensional slab and $\Delta=1$ (isotropic Heisenberg model). 
Red dot: $d=1$, blue square: $d=2$, grey diamond: $d=3$.
$N$ is the linear size of the system. 
In 2D (3D) we consider a square (cubic) lattice with $N\times 5$ ($N\times 5 \times 5$) spins.
In 2D the sender and the receiver are located as in Fig.\ref{fig1}. In 3D they are located in the center
of two opposite faces of the cube, at distance $a$ from the central spin.
Fidelity increases with the the dimensionality of the system for sufficiently large system sizes. 
}\label{fig7}
\end{figure}

\bibliography{biblio.bib}

\end{document}